\documentclass{emulateapj}
\usepackage{amsfonts, color, hyperref, epsfig,textcomp}
\usepackage{apjfonts, natbib}

\newcommand{\chisq}{\ensuremath{\chi^2}}
\newcommand{\aox}{\ensuremath{\alpha_{\rm{ox}}}}
\newcommand{\flamb}{erg~s$^{-1}$~cm$^{-2}$ \AA$^{-1}$}
\newcommand{\flux}{erg~s$^{-1}$~cm$^{-2}$}
\newcommand{\lum}{erg~s\ensuremath{^{-1}}}
\newcommand{\lbol}{\ensuremath{L\mathrm{_{bol}}}}
\newcommand{\ledd}{\ensuremath{L\mathrm{_{Edd}}}}
\newcommand{\lratio}{\lbol/\ledd}
\newcommand{\oiii}{[O\,{\footnotesize III}]}
\newcommand{\ha}{H\ensuremath{\alpha}}
\newcommand{\hb}{H\ensuremath{\beta}}
\newcommand{\nii}{[N\,{\footnotesize II}]}

\newcommand{\msun}{\ensuremath{M_{\odot}}}

\newcommand{\mbh}{\ensuremath{M_\mathrm{BH}}}
\newcommand{\mnsc}{\ensuremath{M_\mathrm{NSC}}}
\newcommand{\mgal}{\ensuremath{M_\mathrm{gal}}}
\newcommand{\hst}{\emph{HST}}
\newcommand{\xmm}{\emph{XMM$-$Newton}}
\newcommand{\xmms}{\emph{XMM}}
\newcommand{\chandra}{\emph{Chandra}}

\newcommand{\sersic}{S\'{e}rsic}
\shorttitle{IMBH in NGC~3319} 
\shortauthors{JIANG ET AL.}
\begin{document}

\title{Discovery of An Active Intermediate-Mass Black Hole Candidate in the Barred Bulgeless Galaxy NGC~3319}
\author{Ning~Jiang\altaffilmark{1,2},
Tinggui~Wang\altaffilmark{1,2}, 
Hongyan~Zhou\altaffilmark{1,3}, 
Xinwen~Shu\altaffilmark{4}, 
Chenwei~Yang\altaffilmark{3},
Liming~Dou\altaffilmark{5},
Luming~Sun\altaffilmark{1,2},
Xiaobo~Dong\altaffilmark{6},
Shaoshao~Wang\altaffilmark{4},
Huan~Yang\altaffilmark{7}
}
\altaffiltext{1}{Key laboratory for Research in Galaxies and Cosmology,
Department of Astronomy, University of Science and Technology of China,
Chinese Academy of Sciences, Hefei, Anhui 230026, China; ~jnac@ustc.edu.cn}
\altaffiltext{2}{School of Astronomy and Space Sciences,
University of Science and Technology of China, Hefei, Anhui 230026; ~twang@ustc.edu.cn}
\altaffiltext{3}{Polar Research Institute of China, 451 Jinqiao Road, Shanghai, 200136, China; ~zhouhongyan@pric.org.cn}
\altaffiltext{4}{Department of Physics, Anhui Normal University, Wuhu, Anhui, 241000, China}
\altaffiltext{5}{Center for Astrophysics, Guangzhou University, Guangzhou, 510006, China}
\altaffiltext{6}{Yunnan Observatories, Chinese Academy of Sciences,
Kunming, Yunnan 650011, China; Key Laboratory for the Structure and Evolution of
Celestial Objects, Chinese Academy of Sciences, Kunming, Yunnan 650011, China}
\altaffiltext{7}{Las Campanas Observatory, Carnegie Institution for Science, Casilla 601, La Serena, Chile}

\begin{abstract}
We report the discovery of an active intermediate-mass black hole (IMBH) candidate
in the center of nearby barred bulgeless galaxy NGC~3319.
The point X-ray source revealed by archival \chandra\ and \xmm\ observations
is spatially coincident with the optical and UV galactic nuclei from
\emph{Hubble Space Telescope} observations. The spectral energy distribution
derived from the unresolved X-ray and UV-optical flux is comparable with
active galactic nuclei (AGNs) rather than ultra-luminous X-ray sources, 
although its bolometric luminosity is only $3.6\times10^{40}~\rm\lum$.  
Assuming an Eddington ratio range between 0.001 and 1, the black hole mass (\mbh) 
will be located at $3\times10^2-3\times10^5$~\msun, placing it in the 
so-called IMBH regime and could be the one of the lowest reported so far. 
Estimates from other approaches (e.g., fundamental plane, X-ray variability) also 
suggest \mbh$\lesssim10^5$~\msun.
Similar to other BHs in bulgeless galaxies, the discovered IMBH 
resides in a nuclear star cluster with mass of $\sim6\times10^6$~\msun.
The detection of such a low-mass BH offers us an ideal chance to study the formation 
and early growth of SMBH seeds, which may result from the bar-driven inflow in
late-type galaxies with a prominent bar such as NGC~3319.
\end{abstract}


\section{Introduction}

It's widely believed that supermassive black holes (SMBHs) with masses of
$10^{6-10}$~\msun\ are present in most (possibly all) galaxies with massive bulges,
and the BH mass correlates tightly with various classical bulge properties
(see Kormendy \& Ho 2013 for a review). These black holes must be grown from
much smaller 'seeds'. Intermediate-mass black holes (IMBHs) in the nearby galaxies,
with masses in the range of $10^{2-6}$~\msun\, provide important clues to
the mass and abundance of these seeds in the early universe
(see review by Greene 2012). IMBHs also bridge the gap between SMBHs in the
galactic nuclei and stellar black holes in binaries. Apparently, IMBHs are
much more difficult to find because the radii of their gravitational influence
are too small to be resolved spatially even in nearby galaxies. 
An alternative approach is to search for dwarf
active galactic nuclei (AGNs), and hitherto discovers hundreds of candidates with
BH masses of $10^{5-6}$ \msun\ (Greene \& Ho 2007; Dong et al. 2012;
Reines et al. 2013; Lemons et al. 2015; Pardo et al. 2016; Mezcua et al. 2016, 2018;
Liu et al. 2018).

In the hierarchical framework of galaxy formation and evolution, the correlation 
between SMBHs and galactic bulges are regulated by the major merger processes. 
A related issue is the observational census of black holes in late-type spirals and
particularly whether supermassive black holes can form in galaxies that lack bulges,
which haven't undergone violent evolution by major merger.
The stellar dynamical constraints on the presence of the BH in local group 
bulgeless galaxy M~33 shows that it does not contain a SMBH with an extremely tight 
upper limit of $\mbh \lesssim 1500-3000$ \msun\ (Gebhardt et al. 2001). 
Again, evidence that BHs can occur in at least some very late-type disk galaxies
comes from the detection of a small number of AGNs in Scd and Sd-type spirals.
The first and most well-studied BH evidence in bulgeless galaxy is NGC~4395
(Filippenko \& Sargent 1989; Filippenko \& Ho 2003), which is a dwarf Sdm galaxy with 
hallmark signatures of a type~1 AGN (Filippenko \& Sargent 1989; Filippenko \& Ho 2003)
such as broad emission lines (Filippenko \& Sargent 1989) and
rapid X-ray variability (Iwasawa et al. 2000). 
A ultraviolet (UV) reverberation-mapping measurement gave
\mbh~$=~(3.6~\pm~1.1)~\times~10^5$~\msun~(Peterson et al. 2005), which is 
verified by recent direct dynamical measurement (den Brok et al. 2015).
More evidences in other bulgeless galaxies are reported in recent years (e.g.,
Shields et al. 2008; Satyapal et al. 2007, 2016; Secrest et al. 2012).
Instead of bulges, nature seems usually to make much more compact nuclear star clusters
(NSCs) at the centers of very late-type spirals (e.g., B{\"o}ker et al. 2002),
with the size of globular clusters but ten times higher in the mass.
It's notable that the discovery of an AGN in a bulgeless galaxy is nearly
always accompanied with a NSC (e.g., Filippenko \& Ho 2003; Shields et al. 2008; 
Barth et al. 2009; Secrest et al.2013).
However, the connection between NSCs and BHs, their link to galaxy formation
and evolution, are still poorly understood.

\begin{figure*}
\centering
\includegraphics[width=14cm]{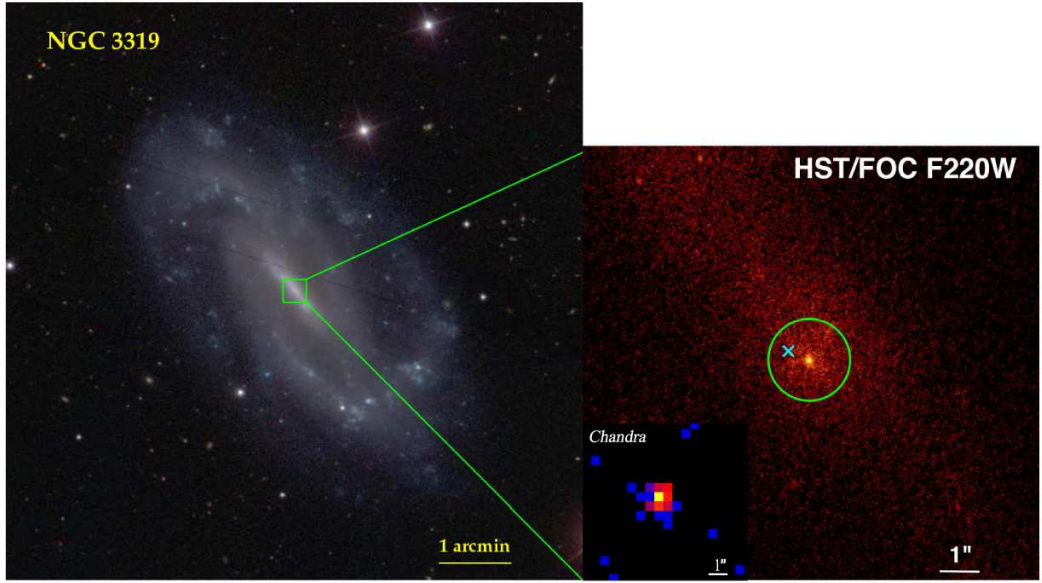}
\caption{
Left: SDSS $gri$ composited image of NGC~3319; right: \hst/FOC UV (F220W) image
of the central part,
the bright compact center is denoted by a green circle with radius of 1\arcsec;
the cyan cross is the point source position detected by \chandra, whose
image is shown in the bottom left inset.
All images are oriented so that north is up, and east is to the left.}
\label{img}
\end{figure*}

\begin{figure*}
\centering
\includegraphics[width=15cm]{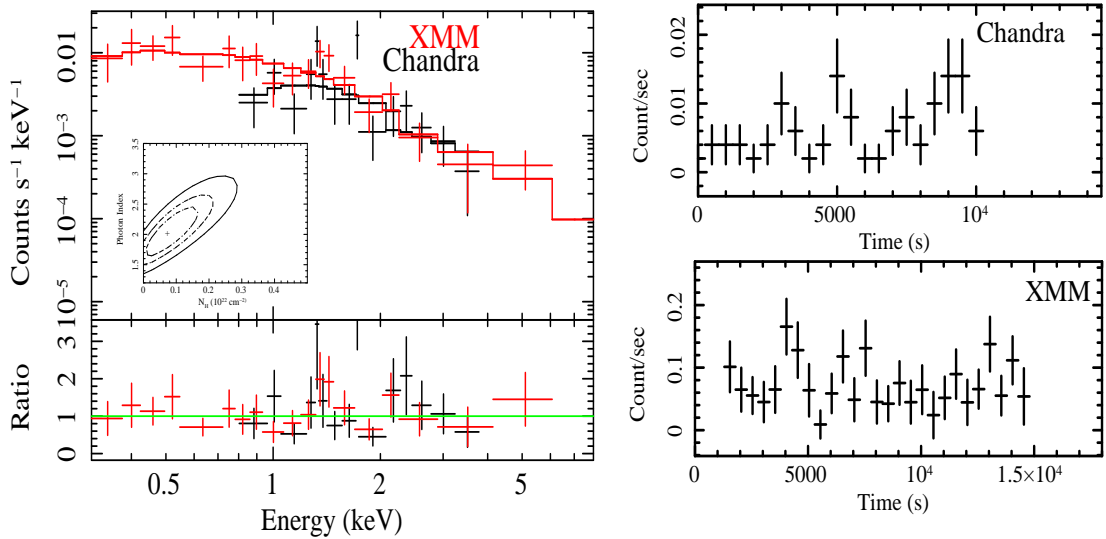}
\caption{
{\it Left:} \chandra\ and \xmms\ spectrum of NGC 3319, along with the best-fit models as
described in Section 2.1.
The inset panel shows the $\chi^2$ contours of the photon index vs N$_{H}$ at the 68\%,
90\% and 99\% confidence levels
(from the inside out).
The data to model ratio showing in the lower panel.
{\it Right:} \chandra\ and \xmms\ light curve of NGC 3319 with a binsize of 500s.
}
\label{xray}
\end{figure*}

In this paper, we report the discovery of an IMBH candidate in the 
center of NGC~3319, which is a nearby barred bulgeless galaxy  
(see its SDSS image in the left panel of Figure~\ref{img}).
This target was firstly noted by our cross-match between \xmm\ (\xmms\ for short) 
serendipitous catalog and nearby late-type galaxies. 
Recent \chandra\ observation pinpoints the X-ray source to the galaxy center.
We adopt the luminosity distance (14.3~Mpc) determined from Cepheid variable stars 
(Sakai et al. 1999). The physical scale at the distance is 69~pc/\arcsec.

\section{Data and Analysis}

\begin{figure*}
\centering
\includegraphics[width=15cm, angle=0]{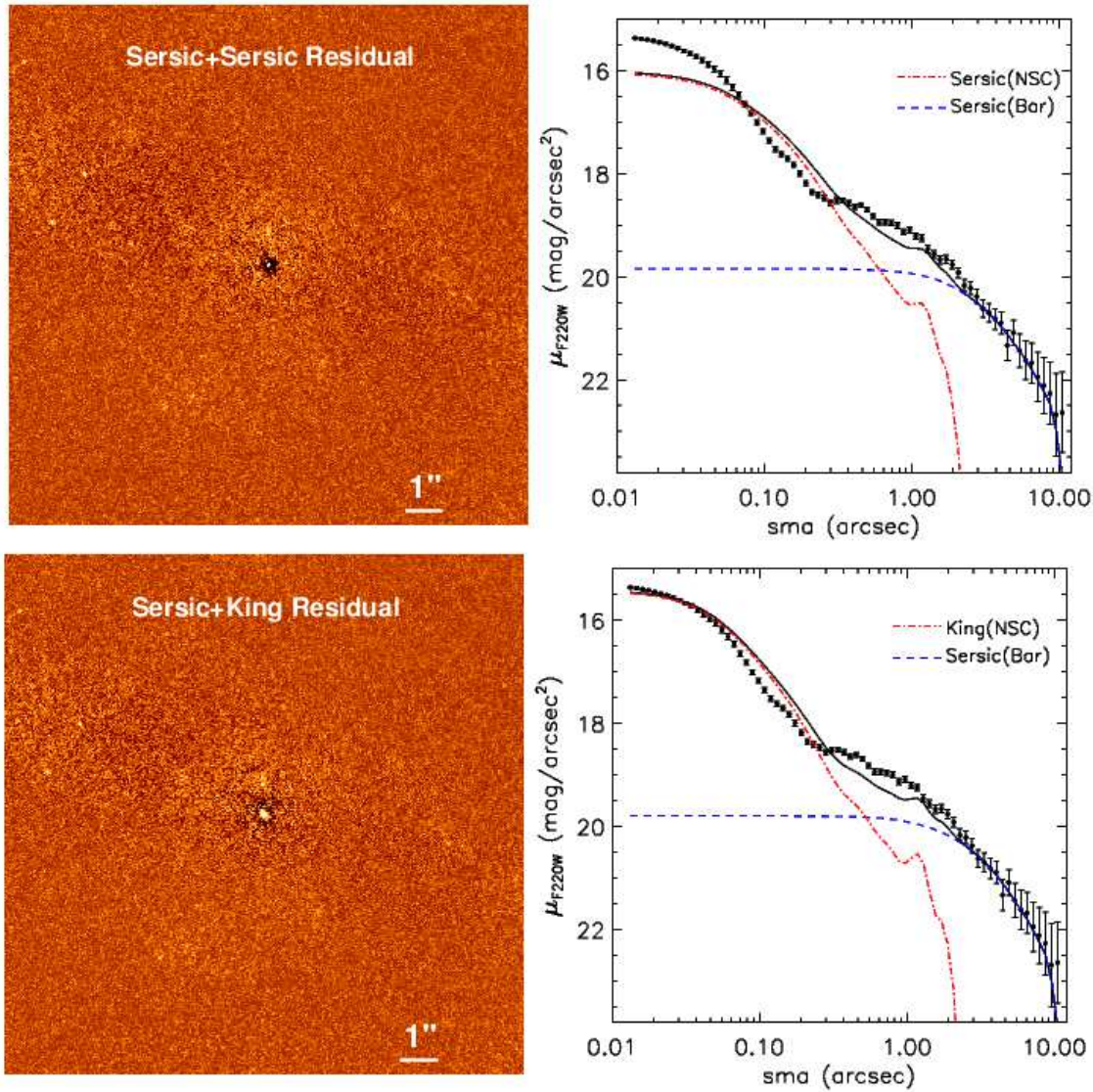}
\caption{
The GALFIT fitting of \hst\ UV (FOC/F220W) image of NGC~3319 without an AGN component.
The fitting residuals are presented in the left and the 1-D surface brightness profiles
are shown in the right panel.
We have tried \sersic\ (top panels) and King (bottom panels) models for the NSC, respectively.
Both fitting schemes lead to large systematical mismatch in the central part.
}
\label{resid}
\end{figure*}

\subsection{X-ray Observations}

NGC~3319 was observed by \chandra\ on 2017 Jan 16 (observation ID 19350, PI: McHardy)
for an exposure of 10 ks, using the back-illuminated chips of the Advanced CCD Imaging
Spectrometer (ACIS-S). The galaxy was placed at the aimpoint of the S3 chip.
The data were processed with the CIAO (version 4.9) and CALDB
(version 4.7.7), following standard criteria.
New level 2 event files were created using the {\tt chandra\_repro} script in CIAO.
We ran the automated point-source detection tool {\tt wavdetect}
on 0.3--8 keV images of the S3 chip and found a point-like X-ray source which 
is spatially coincident with the optical nucleus of NGC~3319 (see detail in Section 3.1).
Taking it as the source position, we extracted the source spectrum
in the 0.3-7.0 keV range from a circular region with a radius of 2\arcsec.5.
The background spectrum was estimated in an annulus region centered on the
source position, with an inner radius of 4\arcsec\ and outer radius of 7\arcsec, respectively.
We also checked that no background flaring events occurred during the observation.
There are totally 61 net counts detected in the energy band of 0.3--7 keV,
but most (50/61) are in the 1-5 keV.
Utilizing the same extraction regions as used for the spectral analysis, 
a background subtracted light curve was created using {\tt dmextract} tool in CIAO.
We have also noted that NGC~3319 is serendipitously detected by \xmms\ 
as a point source on 2004 Aug 24 with an effective exposure time of 10~ks
and the position is consistent with \chandra\ within $\sim0\arcsec.5$.
We extracted counts from a circle with radius of 40\arcsec\ centered on the source, 
and the background is estimated from three circles with 
radii $30-40\arcsec$ around the source.
The extracted X-ray spectrum and light curves are presented in Figure~\ref{xray}.

We have tried to perform joint fit to the \xmms\ and \chandra\ spectrum 
simultaneously. Due to the small number of source counts, the spectrum was 
rebinned to at least 5 counts in each energy bin for \xmms\ and 2 counts for \chandra.
The spectral fitting was performed using {\sc XSPEC} with Cash-statistic in 
the minimization instead of $\chi^2$. We fitted the spectrum with an 
absorbed powerlaw model ($phabs*zphabs*zpowerlaw$), in which $phabs$ is 
the Galactic absorption that is fixed in the fits. 
The model gives a photon index $\Gamma = 2.02$~(1.60-2.50 in 90\% confidence ranges)
and an intrinsic absorption column density $N_{\rm H}=0.8$~(0-1.8)$\times10^{21}\rm cm^{-2}$.
The unabsorbed 2--10~keV flux for \chandra\ observation 
is $4.3~(3.4-5.6)\times10^{-14}$~\flux, corresponding to a luminosity of  
$1.0(0.8-1.4)\times10^{39}$~\lum. Note that the \xmms\ luminosity is higher 
by a factor of $\sim1.5$, but consistent with each other within the errors.
We will adopt the \chandra\ result in the analysis below since it has better resolution.

\subsection{HST Observations}

\begin{figure*}
\centering
\includegraphics[width=19cm]{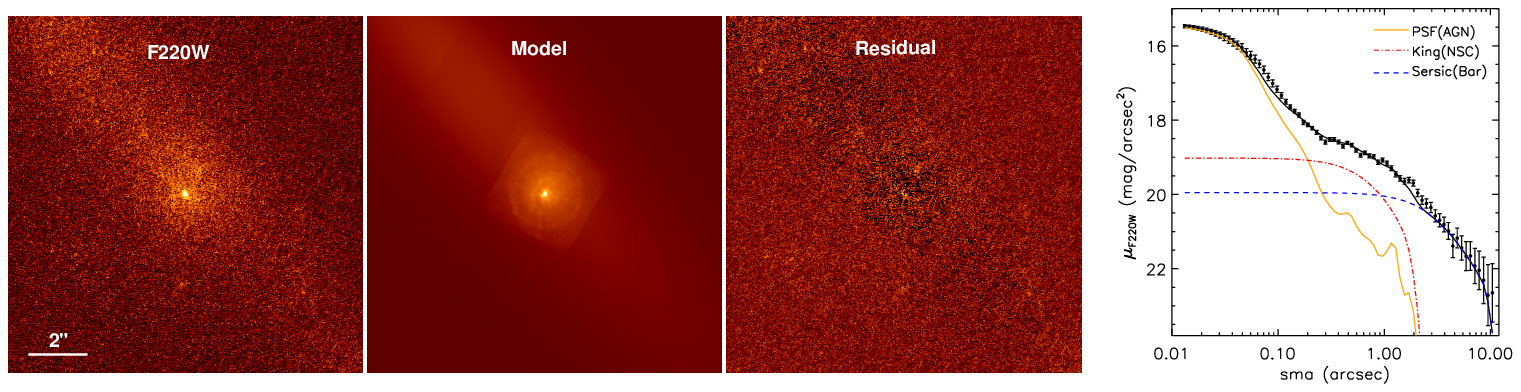}
\includegraphics[width=19cm]{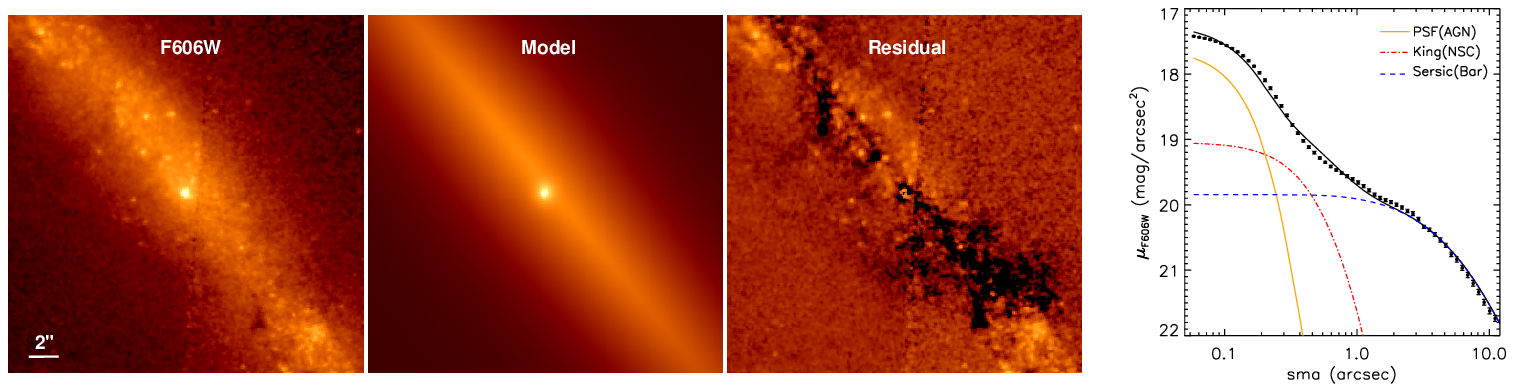}
\caption{
The Galfit decomposition of \hst\ UV (top panel, FOC/F220W) and optical
(bottom panels, WFPC2/F606W) image of NGC~3319.
The PSF, King and \sersic\ components are adopted to represent for the
AGN, NSC and bar flux, respectively.
The left three panels show the original, model and residual images, the right
panel is the 1-D presentation of the decomposition.
}
\label{hst}
\end{figure*}

NGC~3319 has been observed by \emph{Hubble Space Telescope} (\hst) 
Wide Field and Planetary Camera 2 (WFPC2) over a two-month period  
(1997 Nov 11 -- 1998 Jan 3), including thirteen epochs of F555W and four epochs of F814W
exposures. Before that, it has been targeted in the F606W filter on 1994 Oct 19
and in UV (F220W) by Faint Object Camera (FOC) on 1993 Mar 15.

\subsubsection{UV image}

The UV image of NGC~3319 was observed as part of the project to search for 
low-luminosity AGNs that appear as unresolved UV point sources in the nuclei of galaxy
down to resolution $\sim$0\arcsec.05 with pre-COSTAR \hst/FOC (Proposal ID: 4804).
A compact source is embedded in the elongated diffuse emission along with the bar,
which was first noticed as unresolved by Maoz et al. (1996).
To check it and measure the brightness of the UV source, we try to perform 
two-dimensional decomposition using GALFIT (Peng et al. 2002; 2010). 
Precise point spread function (PSF) is of great importance to separate the 
compact central object from the extended starlight. 
However, the pre-COSTAR PSFs change with time since the telescope 
was shrinking fairly rapidly during the early years of the mission 
due to water loss from the carbon fiber telescope body. 
We adopted the stellar PSF taken in 1992 November\footnote{
http://www.stsci.edu/ftp/instrument\_news/FOC/Foc\_tools/psfs/f96\_nov1992.html} 
as an acceptable match to our data.

We begin the fitting with a two-\sersic\ model, in which one represents the compact 
emission and the other for the extended underlying bar.
The free fitting yields an unreasonable high \sersic\ index ($n=19.93$) and large 
effect radius ($r_e$) for the central component.
Hence we tried to fix the $n$ to be 0.5, 1, 2, 3 and 4 respectively and run it again.
The best fitting (with least $\chisq$) corresponds to $n=2$ and $r_e=1\arcsec.68$,  
which has given clean residuals for the outer region yet with significant 
mismatch in the central 2\arcsec\ region (see the top panels of Figure~\ref{resid}).
Then we attempted to replace the central \sersic\ component with a standard empirical 
King profile (King 1966), which is a more popular model to characterize star cluster.
The free fitting crashed immediately because of an unacceptably large 
truncation radius ($r_t$) as well as small core radius ($r_c$), thus we fixed the 
$r_t$ to different values (e.g., 30, 50, 100 pixels).  
The small $r_t$ (30 pixels) case yields similar systematical mismatch with the two-\sersic\
scenario while the large $r_t$ (100 pixels) case fits the outer part much better
except for a deficient filling in the very central part (<0\arcsec.2).
The obvious nuclear excess (see bottom panels of Figure~\ref{resid})
implies an unresolved or extremely compact component.

\begin{deluxetable*}{c|ccccccc}
\tabletypesize{\small}
\tablewidth{0pt}
\tablecaption{GALFIT Decomposition \label{tbl-galfit}}
\tablehead{ \colhead{Band} & \colhead{Component} & \colhead{$m$} & \colhead{$M$}
& \colhead{$n$} & \colhead{$r$ (\arcsec/pc)} & \colhead{$b/a$} & \colhead{$c$} \\
(1) & (2)    & (3)  & (4) & (5) & (6) & (7) & (8)}
\startdata
F220W  & PSF      & 18.70  & $-12.19$ & \nodata   & \nodata   & \nodata & \nodata  \\
       & King     & 18.17  & $-12.72$  & [2] &  0.60/41 & 1.0 & \nodata        \\
       & \sersic\ & 15.01  & $-15.88$  & 0.67  & 8.52/587  & 0.24 & $-0.13$       \\
F606W  & PSF      & 20.14  & $-10.67$  & \nodata   & \nodata   & \nodata   & \nodata \\
       & King     & 18.73  & $-12.08$  & [2]  & 0.45/31   & 0.88 & \nodata  \\
       & \sersic\ & 13.25  & $-17.56$  & 0.93  & 36.8/2537  & 0.13 & $-0.06$
\enddata
\tablecomments{ Col. (1): $HST$ filter.
Col. (2): Components used in the fitting schemes.
Col. (3): The fitted magnitudes, not corrected for Galactic extinction.
Col. (4): The absolute magnitude after Galactic extinction correction.
Col. (5): The index for \sersic\ or $\alpha$ for the King profile.
Col. (6): The effective radius of the \sersic\ component or core radius of King profile,
in units of arcsec and pc, respectively.
Col. (7): Axis ratio.
Col. (8): Diskiness (negative)/boxiness (positive) parameter,
defined in Eqn.~(3) of Peng et al. (2002).
The brackets mean that they are fixed.
The formal errors given by GALFIT are all tiny:
$<0.05$ for magnitude and \sersic\ index, $<0.\arcsec1$ for $r$.}
\label{tbl-galfit}
\end{deluxetable*}

In order to depict the nuclear excess in a more self-consistent way,
we then performed GALFIT decomposition by adding an unresolved PSF component.
We start with the PSF+ King + \sersic\ model.
The fitting has improved a lot (with reduced $\chisq$ changed from 1.12 to 1.07) and yielded 
smooth residuals at both small and large scales (see top panels of figure~\ref{hst}).
Given a zeropoint of 20.64 in the ST magnitude system, 
the fitted magnitudes of the PSF and King component are $18.70\pm0.01$
and $18.17\pm0.03$, corresponding to $(1.33\pm0.03)\times$ 
and $(2.17\pm0.09)\times10^{-16}$~\flamb, respectively.
To test whether the magnitudes are model dependent, we then try the PSF+2-\sersic\ model, 
that is to fit the NSC with a \sersic\ component (e.g., Carson et al. 2015) 
instead of a King profile. The fitting yields slightly larger residuals yet with a similar 
PSF magnitude of $18.66\pm0.01$, which is 0.04 magnitude offset from the King model.
We take the offset as the uncertainty of the PSF component.
The final best-fit parameters are presented in Table~\ref{tbl-galfit}.

We have also tried the PSF model created by Tinytim (Krist 1995), which has 
considered the spherical aberration, and obtained comparable results yet with 
larger residuals in the central part than empirical star PSF.

\subsubsection{Opitcal Image}

The step-by-step decomposition is not trivial for WFPC2 optical images.
First, the resolution in the optical is lower than the UV, particularly 
considering that the center of NGC~3319 is located in the WF4 camera 
(0\arcsec.1/pixel) and thus severely undersampled.
Moreover, the F555W and F814W images are all saturated in the central part,
but the fact itself indicates a bright nucleus.
The F606W image is the only unsaturated optical image because of a shorter exposure time.
For ease of rejecting cosmic rays, the observation was splitted into 
2 equal exposures of 80 seconds each.
The two single exposures are then combined, resulting in the cosmic-ray cleaned image
which is ready for the further GALFIT fitting.
Following our previous work with F606W imaging decomposition (Jiang et al. 2013),
we used the modeled PSF generated by Tinytim.

The same model as the UV band is exploited in the fitting and yields a fairly 
good result (see bottom panel of Figure~\ref{hst}). 
The PSF and NSC magnitudes given by the GALFIT fitting are 
$20.14\pm0.01$ and $18.89\pm0.02$, respectively.

\subsection{SDSS Spectrum}

The center of NGC~3319 was spectroscopically observed by SDSS on 2004 Jan 29 with an
exposure time of 2500~s. 
We fit the spectrum with the BC03 stellar population model (Bruzual \& Charlot 2003) 
using the STARLIGHT code (Cid Fernandes et al. 2005).  
During the fitting, we have masked out all the prominent emission line regions.
The fitted starlight model matches both the continuum and most of the main absorption 
lines well and no extra non-stellar component is needed. 
After subtracting the stellar population model from the raw spectra,
no evident broad emission lines are visible but weak narrow emission lines of 
\hb, \oiii, \ha\ and \nii\ are detectable in the residual spectrum. 
We measure the flux of each emission line with a single Gaussian model.

\begin{figure*}
\centering
\includegraphics[width=15cm]{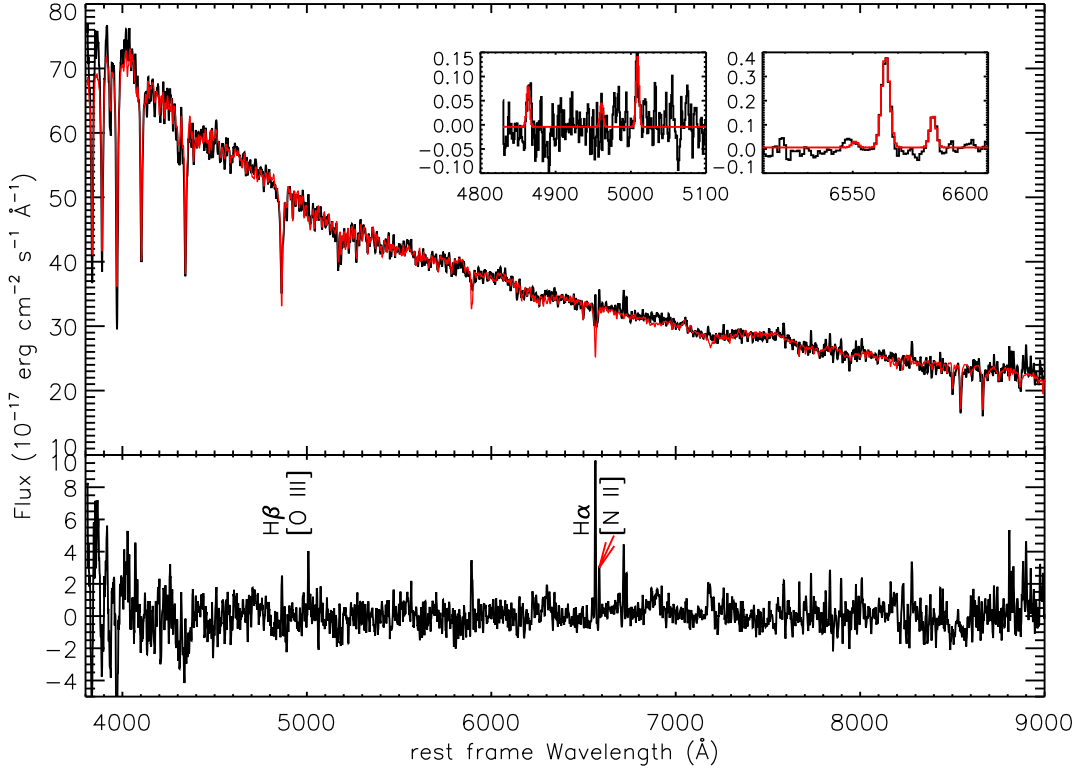}
\caption{
The SDSS spectrum of NGC~3319 (black) with the fitted starlight model plotted in red.
The residual spectrum is shown on the bottom.
The two insets highlight the \hb-\oiii\ and \ha-\nii\ region, respectively.
We fit each emission line with a single Gaussian function.
}
\label{sdss}
\end{figure*}

The BPT line-ratio diagram (Baldwin, Phillips \& Terlevich 1981) are commonly adopted
to discriminate between AGN and SF activity. 
The NGC~3319 is located at the star-forming region of the BPT diagram basing on the fitted 
line fluxes yet with large uncertainty.
The absent AGN signature can be explained by the overwhelming dominance of the stellar light 
in the SDSS spectrum, which was acquired by a fiber with aperture of 3\arcsec.
We have integrated the flux within SDSS fiber from the \hst/WFPC2 F60W image 
and found a magnitude of 17.23, that is 15 times brighter than the fitted AGN component.
We caution that the emission line fluxes are much smaller than the stellar continuum, 
so a small deviation (e.g., a few percent) of the stellar population subtraction 
may lead to significant difference in the residual Balmer lines.
We have also noticed that NGC~3319 has been previously observed by double spectrograph 
on the Hale 5~m telescope and the spectra in the central $2\arcsec\times4\arcsec$ 
region has been classified as uncertain H~II nucleus (Ho et al. 1995, 1997).

The stellar population of NGC~3319 is quite young with an average age of only $130$~Myr 
given by the fitting. If we naively assume that the NSC possesses the same stellar population 
as that revealed by the SDSS spectrum, which has collected substantial starlight outside
of the NSC, the mass of the NSC (\mnsc) is $5.7\times10^6$~\msun.
As a comparison, Georgiev \& B{\"o}ker (2014) has obtained a much more compact 
NSC with $r_e\sim9$~pc and \mnsc$\sim1.6\times10^6$~\msun\ (Georgiev et al. 2016).
The discrepancy can be addressed as they haven't set the PSF component in their fitting, 
which will force the NSC to match the central unresolved component.

\section{The Nature of the Unresolved Nuclear Emission}

\subsection{Coincidence of the X-ray and optical/UV Point Source}

The X-ray luminosity ($\sim10^{39}$~\lum) revealed by \xmm\ and \chandra\ observations
can not tell us immediately the nature of the source, e.g., an ultraluminous 
X-ray source (ULX) powered by X-ray binaries (XRBs) or a low-luminosity AGN.
ULXs are usually defined as point-like sources within the optical extent of a host galaxy 
but away from the nucleus in order to exclude AGNs (see review by Kaaret et al. 2017).
Several different isotropic X-ray luminosity threshold have been used to classify ULXs,
such as the  Eddington limit for a 1.4~\msun\ neutron star (Makishima et al. 2000) 
or simply $1\times10^{39}$~\lum (e.g., Swartz et al. 2011).
First, we explore the possibility of XRBs by virtue of the scaling relations
between the X-ray emission from XRBs and star formation rate (SFR).
Using the recent formalism given by Lehmer et al (2016) and global SFR of NGC~3319
(Zhou et al. 2015), the expected global X-ray luminosity from high-mass XRBs is 
$2.0\times10^{38}$~\lum, that is 5 times lower than the observed value.
If we only care about the bar SFR, the predicted X-ray luminosity would be even lowered by 
one order of magnitude.

Second, the X-ray source position is exactly at or extremely close to the galactic center.
The \chandra\ X-ray source coordinate is RA=10:39:09.446, DEC=+41:41:12.10, 
that is 0.16\arcsec\ away from the optical center given by NED.
The UV center given by \hst/FOC image 
is RA=10:39:09.401, DEc=+41:41:11.91, that's 0\arcsec.54 from the X-ray position 
(see right panel of Figure~\ref{img}) yet still within the pointing error 
due to the guide star position uncertainty (1\arcsec).~\footnote{See the handbook on \hst\ target acquisition in section 5.2.1 of 
http://documents.stsci.edu/hst/proposing/documents/pri\_cy10/primer.pdf}
We have also checked the high-resolution optical center by \hst\ images and found that
they're all consistent with each other when taking into consideration of errors.
Hence, under current astrometry accuracy, the X-ray point source is highly consistent with 
the optical and UV galactic center, that is in disfavour with the ULX scenario since ULXs are 
defined as off-nucleus.
On the other hand, according to the statistics on local ULX number density, ULXs are
detected at rates of one per $3.2\times10^{10}$~\msun, one per $\sim0.5$~\msun~$\rm yr^{-1}$
star formation rate (Swartz et al. 2011). It will predict a chance of $<10^{-4}$
to find a ULX in the circle of \chandra\ resolution.

\subsection{SED: Consistent with High-accretion-rate AGNs}

Once both the unresolved optical-UV and X-ray radiation are convincingly assumed to 
originate from the same source, we can try to explore its nature by 
spectral energy distribution (SED) characteristics.
First of all, ULXs usually don't show or have very weak optical counterparts.
The absolute magnitudes of well-studied ULX optical counterparts are all fainter than
$M_{V}\approx-8$, with a median magnitude of $-6$ (Vinokurov et al. 2018).
Our GALFIT fitting of F606W image (roughly $V$-band) gives a PSF magnitude of -10.67, 
that is significantly more luminous than normal ULXs.
Furthermore, the X-ray to optical flux ratio, defined as 
$\log(f_{\rm X}/f_V) = \log f_{\rm X} + m_V/2.5 + 5.37$ (Maccacaro et al. 1982), 
where $f_{\rm X}$ is the 0.3-3.5 keV observed flux in \flux\ and 
$m_V$ is the visual magnitude, can be used to distinguish AGN, BL Lac objects, 
and X-ray binaries (Stocke et al. 1991).
The calculated $\log(f_{\rm X}/f_V)$ of NGC~3319 is 0.25, that is 
apparently lower than usual ULXs with ratio $>2$ (Tao et al. 2011).
Hence both the position and the bright optical counterpart of the nuclear source are
inconsistent with the ULX scenario. 

The spectral energy distribution (SED) of luminous Seyfert galaxies and quasars 
are well-known characterized by a big blue bump in the optical and UV bands, which is 
thought to be thermal emission from an optically thick accretion disk extending 
to a few gravitational radius (e.g., Shields 1978; Malkan \& Sargent 1982). 
This is also the case for the optical-UV-X-ray SED of NGC~3319 nucleus(see Figure~\ref{sed}).
Such an SED is very close to that of high accretion rate 
AGN systems ($\log(L/L_{Edd})>-1.0$ in Ho (2008)), but very different from those of 
low accretion rate systems in nearby low luminosity AGNs (Ho 1999, 2008).
In addition, the ratio of the optical-to-X-ray flux (\aox) is usually exploited 
by AGN community to specifically the balance of energy coming out in the optical/UV
emerging from the accretion disk as compared with the X-ray
luminosity from the corona (e.g., Tananbaum et al. 1979).
We adopt the universal definition $\aox\equiv-0.3838\log{f_{\rm 2500~\AA}/f_{\rm 2~keV}}$,
where $f_\nu \propto\nu^{\aox}$ is the specific flux.
Hence, it gives $\aox=-1.40\pm0.05$, in which $f_{2500~\AA}$ is derived from the
F220W flux assuming a spectral index -1.5 and the $f_{\rm 2~keV}$ is from \chandra.
The calculated \aox\ has well fallen in the range of AGNs including
IMBHs (Dong et al. 2012b; Baldassare et al. 2017; Liu et al. 2018), and suggests a central BH accreting system.

\subsection{BH Mass Estimate}

The \mbh\ of type~1 AGNs are usually computed by the empirical formula under 
the assumption of virial equilibrium of broad-line region gas (e.g., Greene \& Ho 2005). 
Unfortunately, the SDSS spectrum of NGC~3319 
is totally dominated by starlight and no broad \ha\ or \hb\ line is present. 
To catch sight of the AGN activity (such as potential broad lines) of such a small BH, 
further high-resolution spectra beyond seeing-limited observations 
(e.g., by \hst\ or ground-based telescopes with adaptive optics) are probably required.
Nevertheless, we can still try to estimate the \mbh\ by some other means. 

\begin{figure}
\includegraphics[width=8cm]{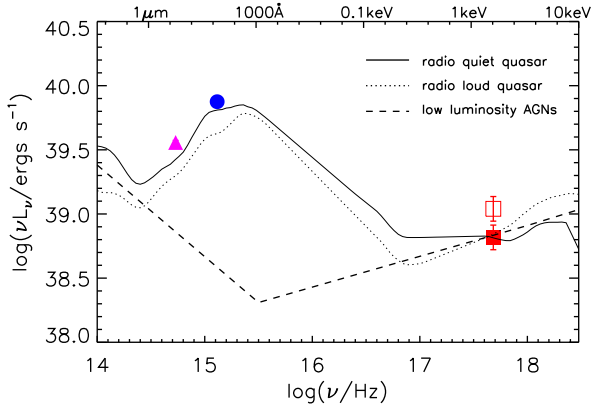}
\centering
\caption{
The X-ray (red filled square for \chandra\ and red open square for \xmms),
UV (blue filled circle) and optical (magenta triangle) luminosity of the unresolved source
in the center of NGC~3319.
The SEDs of median radio quiet (black line) and loud (dotted line) quasar SED
(Elvis et al. 1994) and median low luminosity AGNs
(dashed lines, adapted from Ho et al. 1999) are overplotted.
They are normalized to the 2~keV luminosity.}
\label{sed}
\end{figure}

Despite only a sparse sampling of the AGN SED of NGC~3319 is available 
(see Figure~\ref{sed}), it should still offer a more reliable measurement of the 
bolometric luminosity (\lbol) of the system than any estimate based on a single band.  
We have first scaled the median radio-quiet quasar SED of Elvis et al. (1994)
to the \chandra\ X-ray luminosity and then integrated the whole SED, the resulted \lbol\ is   
$\sim3.6\times10^{40}$~\lum. If we scale the quasar SED to the optical or UV luminosity, 
the difference of the \lbol\ is not greater than 0.1 dex.
Interestingly, the \lbol\ of NGC~3319 is similar to NGC~4395 (Peterson et al. 2005) yet
their SEDs are clearly differentiated with each other.
The SED of NGC~4395 shows no big blue bump with $\alpha_{\rm ox}=-0.97$ and its 
\lratio\ is estimated to be down to $1.2\times10^{-3}$ 
(Moran et al. 1999; Dewangan et al. 2008).
In contrast, the SED of NGC~3319 agrees with a typical quasar. 
Besides, the \lratio\ is found to be moderately correlated with hard X-ray 
photon index ($\Gamma$) in radio-quiet AGNs, using the correlation 
given by Shemmer et al. (2008), we got a \lratio\ of 0.26 yet with an error as
high as one dex. Assuming an Eddington ratio of 0.1, the \mbh\ of NGC~3319 will be 
$3\times10^3\msun$, in the regime of IMBH.
As a conservative estimate of uncertainty, allowing the 
Eddington ratio in the range of 0.001 to 1.0, we estimate the BH 
mass in the range of $3\times10^2$ to $3\times10^5~$\msun.

It is widely known that in AGNs and BH X-ray binaries, there is a tight correlation 
among their radio luminosity ($L_R$), X-ray luminosity ($L_X$) and \mbh, 
that is the so-called "fundamental plane" (FP) of BH activity.
Therefor, the FP could be adopted as an alternative to derive \mbh\ when both 
the X-ray and radio luminosity are available.
The recent high-resolution ($\leq0.2\arcsec$) 1.5~GHz radio image of NGC~3319 
yields non-detection with an upper-limit core luminosity of $10^{34.84}$~\lum\ 
(Baldi et al. 2018).
Taking advantage of the FP embodying low-mass BHs (Qian et al. 2018),
the \mbh\ in NGC~3319 would be $\leq10^{5}$~\msun, that is consistent with
the result above.

We have also tried to estimate the \mbh\ by X-ray variability.
The excess variances of \chandra\ and \xmms\ light curves
are $\sigma^2_{\rm rms}=0.093\pm0.088$ and $\sigma^2_{\rm rms}=0.001$ 
(with 1-$\sigma$ upper limit 0.067), respectively.
We note that the value measured from the \chandra\ is broadly consistent
with that from the \xmms\ within errors.
Although with a large scatter, we estimated the \mbh\ from
the $\sigma^2_{\rm rms}$ from the \chandra\ data, using the relation derived by
Pan et al. (2015) for low-mass AGNs, yielding $\mbh\sim1.5\times10^5$~\msun.
In one word, the \mbh\ of NGC~3319 is well constrained to be at the low end of 
the central BHs, likely $\lesssim10^5$~\mbh\ basing on various means above 
even if the precise mass is difficult to measure for the time being.

\section{Discussions}

\subsection{The Uniqueness of the IMBH in NGC~3319}

Our rough mass estimation of the IMBH in NGC~3319 may be down to $\sim10^3$~\msun,
that is possibly lower than any other known BHs found in the center of galaxies
(e.g., Secrest et al. 2012; Baldassare et al. 2015),
promoting the accumulated IMBH population a step closer to the stellar mass BHs. 
The discovery of such a low-mass BH is extremely rare and its success is 
owing to the particularities of NGC~3319 by comparing with other well-known IMBHs 
(see a summary of their properties in Table~\ref{tb_imbh}).

\begin{deluxetable*}{l|cccccccc}
\tabletypesize{\scriptsize}
\tablecaption{Comparison with other well-known IMBHs}
\tablewidth{0pt}
\tablehead{
\colhead{Galaxy}   &
\colhead{Distance} &
\colhead{$L_{\rm 2-10keV}$}   &
\colhead{\lbol}   &
\colhead{\aox}   &
\colhead{\lratio}   &
\colhead{\mbh}  &
\colhead{Host type} &
\colhead{Reference}  \\
(1) & (2) & (3) & (4) & (5) & (6) & (7) & (8) & (9) }
\startdata
NGC~4395 & 4.2   & $2.3\times10^{40}$ & $5\times10^{40}$       & 0.97 & $1.2\times10^{-3}$ & $4^{+8}_{-3}\times10^5$ & dwarf irregular      & Filippenko \& Ho (2003) \\
POX~52   & 93    & $4.1\times10^{41}$ & $1.3\times10^{43}$     & 1.44 & 0.2-0.5            & (2.2-4.2)$\times10^5$   & dwarf elliptical     & Barth et al. (2004) \\
Henize~2-10 & 9  & $2.7\times10^{39}$ & $2.7\times10^{40}$ & ... & $10^{-4}$ & $\sim2\times10^6$~\msun & dwarf starburst & Reines et al. (2011) \\
UM~625   & 109   & $6.5\times10^{40}$ & (0.5-3)$\times10^{43}$ & 1.72 & 0.02-0.15          & $1.6\times10^6$         & Pseudobulge S0 & Jiang et al. (2013) \\
RGG~118  & 106   & $4.0\times10^{39}$ & $4\times10^{40}$       & ...  & $\sim0.01$         & $\sim5\times10^4$       & dwarf disk           & Baldassare et al. (2015) \\
NGC~4178 & 16.8  & $8.6\times10^{39}$ & $9.2\times10^{42}$     & ...  & $>0.2$             & (1-10)$\times10^4$      & barred bulgeless     & Secrest et al. (2012) \\
NGC~3319 & 14.3  & $1.0\times10^{39}$ & $3.6\times10^{40}$       & 1.40 & $\sim0.1$          & $3(0.3-300)\times10^3$       & barred bulgeless     & This work
\enddata
\tablecomments{
Column~(1): object Name.
Column~(2): luminosity distance in unit of Mpc.
Column~(3): X-ray luminosity integrated from 2 to 10 keV.
Column~(4): bolometric luminosity in unit of \lum.
Column~(5): \aox.
Column~(6): Eddington ratio.
Column~(7): estimated \mbh\ in unit of \msun.
Column~(8): host galaxy type.
Column~(9): literature which reports the discovery of the IMBH.
For the two extensively-studied IMBH prototypes NGC~4395 and POX~52, we
have adopted the most updated parameters. For instance, the \mbh\ of NGC~4395
is measured from the gas dynamic modeling (den Brok et al. 2015).
}
\label{tb_imbh}
\end{deluxetable*}

First, NGC~3319 is one of the most nearby IMBHs.
The distance effect operates not only in the brightness but also
the physical size resolving ability.
Even at \hst\ resolution level, the structural decomposition of 
AGN and starlight is not an easy job for more distant sources, 
particularly when considering cases of co-existence of BH and NSCs (e.g., Shu et al. 2017).
Only with the aid of \hst\ UV and \chandra\ X-ray imaging,
which have the best resolution current available, the SED of the unresolved 
AGN component is obtainable and then places strong constraints of its accretion rate.
Hence, our discovery also implies that a joint UV/X-ray survey of late-type galaxies
may be very efficient to detect low-mass BHs. Comparing with pure X-ray survey
(e.g., She et al. 2017), the added UV band can help us diagnose its accretion state
and the ULX possibility.
Second, the \lbol\ of NGC~3319 is the faintest among all sources in question.
Though it's comparable with NGC~4395, however, its seemingly higher \lratio\
results in a lower \mbh. RGG~118 is the closest source to NGC~3319 in terms of 
\lbol\ and \lratio, whose \mbh\ is probably the lowest derived from the broad \ha\ line
(Baldassare et al. 2015).
Last but not least, the barred bulgeless host of NGC~3319 is also very distinguishing.
The most analogous object before is NGC~4178, which also contains a NSC 
and is located at a similar distance, but with a \lbol\ two orders of magnitudes higher.
The absence of a notable bulge also reduce the starlight contamination to 
recognize the signal from AGN.

As a brief summary, the BH in NGC~3319 is a unique IMBH in terms of its close distance,
low luminosity, high accretion rate, bulgeless host, and most interestingly the small
central BH. Further observations are highly encouraged to get a more precise \mbh.

\subsection{Implications for the Formation and Growth of IMBHs}

Galactic nuclei typically host either a NSC (prevalent in late-type galaxies) 
or a SMBH (common in early-type galaxies), among which the most intriguing observations 
is the coexistence of NSC and SMBH in some galaxies.
Ferrarese et al. (2006) has introduced the term central massive object (CMO)
to unify the BH and NSC and suggest that the formation and evolution of both 
types of central mass concentration may be linked by similar physical processes.
As we have mentioned in the introduction, the central BH discovered in late-type 
galaxies are often associated with a NSC, that's exactly the case of NSC~3319.
Such systems are undoubtedly very valuable for us to probe the 
formation of seed BH and the NSC.

As a well-known rule of thumb, we know that the mass ratio between SMBHs and classical
bulges are $\sim$0.2\% (e.g., Marconi \& Hunt 2003, see a more comprehensive review 
in Kormendy \& Ho 2013). 
Assuming a Eddington rate of 0.1, we get \mbh$\sim3\times10^3$~\msun, that is $0.05\%$ 
of the \mnsc, slighter lower than the relation found for classical bulges. 
On the other hand, it's proposed that similar scaling relations are also hold 
between \mnsc\ and their host galaxy stellar mass (\mgal) albeit it's hotly debated 
what is the physical mechanism setting the
correlation (e.g., Ferrarese et al. 2006; Georgiev et al. 2016).
Given the \mgal=$3.1\times10^9$~\msun (Georgiev et al. 2016), 
the \mnsc\ is only 0.18\% of the \mgal, fairly consistent with classical bulges.
As a comparison, the \mbh/\mnsc\ and \mnsc/\mgal\ ratios are 16\% and 0.26\% in NGC~4395 
respectively, which contains a relatively larger BH and yet comparable NSC mass. 
This may suggest that the BH and NSC in NGC~3319 are lying in the very 
early stage of growth, even earlier than the IMBH prototype object NGC~4395.

Another striking morphology feature of NGC~3319 is the strong bar with a length 
about $100\arcsec$~(7 kpc).
As a typical morphology feature of non-axisymmetric potential,
galactic bar can play an important role in the secular evolution of disk galaxies,
that are capable of driving galactic-scale gas down to approximately pc scales
(e.g., Shlosman et al. 1989; Friedli \& Benz 1993; Wang et al. 2012). 
The resulting gas reservoir in the galaxy centers may serve to feed the BH and thus 
bars have been suspected to be viable mechanism to trigger AGN activity in spiral galaxies
(e.g., Shlosman et al. 1990; Sakamoto et al. 1999; Jogee et al. 2005)
although many statistics on basis of larger sample and with selection effect
accounted for show that the AGN activity is little affected by the presence of 
large-scale bars (e.g., Hao et al. 2009; Lee et al. 2012; Goulding et al. 2017).

The hydrodynamical simulations don't only show that a bar is indeed able to transfer 
gas to the galactic center and form a central mass concentration, but also suggest that 
a seed SMBH could be created (e.g., Sellwood \& Moore 1999; Fanali et al. 2015). 
Li et al (2017) has revisited the model and argue that a NSC and perhaps a massive BH
could form from the concentrated gas of low angular momentum 
during the bar-driven gas inflow, which will stall at a nuclear ring 
when the mass of the central object exceeds 1\% of the disk mass.
NGC~3319 may represent such a typical bar-driven growth scenario of SMBH seeds. 
According to Kormendy \& Kennicutt (2004), a pseudobulge will form soon 
in this galaxy as a result of secular evolution. 
However, our analysis of SDSS spectrum suggests that
the bar was experiencing a starburst in about $\sim$130 Myr from now, but
the major star formation has ceased as there are only weak \ha\ emission lines
from HII regions. Therefore, the formation of the pseudobulge must go
through with stellar dynamic processes rather than in the process of the
formation of new stars. The coincidence of the \mbh/\mnsc\ and \mnsc/\mgal\ ratios
to that of classical bulge suggests perhaps that
NSC is the progenitor of the pseudobulge although the physical process needs to be
further understood.

\acknowledgements

We thank for the anonymous expert referee for many constructive comments and suggestions.
We thank STScI help desk for the questions about \hst\ pointing accuracy and FOC PSFs.
N.J. thanks the valuable comments from Juntai Shen and useful discussion with Mouyuan Sun. 
This work is supported by the National Basic Research Program of China
(grant No. 2015CB857005), NSFC (116203021, 11421303, 11573001, 11733001),
Joint Research Fund in Astronomy (U1431229, U1731104) under cooperative agreement
between the NSFC and the CAS and the Fundamental Research
Funds for the Central Universities.
This research has made use of the NASA/IPAC Extragalactic Database
(NED), which is operated by the Jet Propulsion Laboratory,
California Institute of Technology, under contract with the
National Aeronautics and Space Administration.

\acknowledgements

\end{document}